\documentclass[aps,pra,twocolumn,letterpaper,superscriptaddress,10pt]{revtex4-2}

\usepackage{amssymb,amsthm,amsmath,amsfonts}
\usepackage{graphicx,mathptmx}
\usepackage[pdftex,dvipsnames,usenames]{xcolor}
\usepackage[colorlinks=true,urlcolor=blue,citecolor=blue,linkcolor=blue]{hyperref}
\usepackage{bbm}
\usepackage[shortlabels]{enumitem}
\usepackage{soul}

\begin{document}
\title{Entanglement of particles versus entanglement of fields: independent quantum resources}

\author{Jan Sperling}\email{jan.sperling@uni-paderborn.de}
	\affiliation{Theoretical Quantum Science, Institute for Photonic Quantum Systems (PhoQS), Paderborn University, Warburger Stra\ss{}e 100, 33098 Paderborn, Germany}

\author{Elizabeth Agudelo}\email{elizabeth.agudelo@tuwien.ac.at}
	\affiliation{Atominstitut, Technische Universit\"at Wien, Stadionallee 2, 1020 Vienna \& Institute for Quantum Optics and Quantum Information — IQOQI Vienna, Austrian Academy of Sciences, Boltzmanngasse 3, 1090 Vienna, Austria}

\begin{abstract}
	Nature allows one to explore a manifold of remarkable quantum effects.
	Most prominently, quantum entanglement can be observed in many-particle systems, between multiple quantized fields, and in hybrid combinations thereof.
	This diversity, however, also leads to contradicting conclusions about what truly constitutes entanglement in any given physical scenario.
	By explicitly allowing various perspectives, we rigorously consider different notions of entanglement in the context of first and second quantization.
	By providing instructive examples, we show that particle entanglement and field entanglement are actually distinct phenomena that can occur and be observed independently of each other.
    This conclusion not only affects our fundamental understanding but has direct implications for quantum technology which can harness those independent forms of entanglement in practical scenarios.
\end{abstract}
\maketitle

\section{Introduction}

    Our modern capacity to manipulate and control single quantum systems opened the door to previously inconceivable applications \cite{NC10,DM03}.
    Nowadays, we are pursuing communication, simulation, computation, and sensing in a manner that surpasses their classical counterparts by {harnessing} quantum effects and quantum correlations. 
    More generally, quantum theory delivered a whole new framework for making sense of certain natural phenomena and gave birth to a plethora of different quantum effects that have been explored to advance our fundamental understanding and technological developments.

	Undoubtedly, one of the most fundamental and most widely recognized quantum phenomena is entanglement \cite{HHHH09}.
	Indeed, any generic composite system has the potential to be in a state that exhibits this form of quantum correlation, and many methods exist to detect and characterize entanglement \cite{PV07,GT09,FVMH19}.
	While other exciting forms of quantum correlations exist \cite{FP12,ASW13,KASSVSH21}, entanglement is an archetypal example for how composite quantum systems and their quantum interference properties are in opposition to a classical picture of nature \cite{EPR35,S35,S36}, occurring also between different degrees of freedom in single particles \cite{AMMPP20}.
	Quantum protocols, such as quantum teleportation \cite{BBCJPW93} and superdense coding \cite{BW92}, exploit the nonlocal aspect of entanglement to establish quantum communication between distant parties, which has been proven by recent long-range \cite{NBBBU22} and ground-satellite \cite{Retal17} experiments.
	Furthermore, entanglement was explored in various many-body systems \cite{AFOV08}, ranging from different elementary particles---most notably, photons---to macroscopic structures \cite{S02,HHRBCCKRRSBGDB05,TPOMBTSAZP21}.
	Besides particles, entanglement of quantized fields is the backbone of continuous-variable quantum science \cite{Z02,V03,BL05}.

	Since only a very few requirements have to be satisfied for the emergence of entanglement, we see a richness and variety of entangled states second to none.
	Consider the Bell states---they are the prime example of two entangled qubits;
	but multilevel systems offer even richer high-dimensional entanglement \cite{EKZ20}, featuring larger encoding alphabets than two-level systems.
	Besides the dimensions of the individual subsystems, the number of them further boosts the {variety} of quantum correlations, which applies to theory \cite{HV13,LM14,VSK13,ASI04} and experiments \cite{YUASKSYYMF13,CMP14,GSVCRTF15,GSVCRTF16} alike.
	For instance, already in three qubits there exist two unique classes of states with different entanglement structures, the W and GHZ states \cite{DVC00}.
	And the number of distinct forms of entanglement becomes exceedingly complex {from here on}, e.g., {highly multipartite} graph \cite{HWB04} and hypergraph \cite{RHBM13} states.
	Eventually, in very large systems, macroscopic entanglement can be analyzed \cite{V08,TPKJM16}, even encompassing systems with undefined numbers of subsystems \cite{SW17}, systems with large masses \cite{MRSDC08}, and micro-macro transitions \cite{RSS11}.
    Entanglement is a prime resource for quantum information applications \cite{CH16} and connects to other fundamental phenomena, such as steering and nonlocality \cite{WJD07}.
    It also plays a role in quantum metrology \cite{PS09,T12,HLKSWWPS12,M13}, which, however, is a topic of ongoing debates \cite{BABFMMP18}.

	The purely mathematical structure of entanglement theory is certainly not the most abstract one.
	Nonetheless, there are often debates about what can be considered proper physical entanglement and what bears only a formal resemblance.
	One example is the notion of classical entanglement, where the nonproduct form of classical fields is considered to be an analog for entanglement \cite{S98}, despite not involving any quantum states at all \cite{KB15}.
	For quantum fields, the choice of the mode basis can alter between mode entanglement and disentanglement without actually changing the physical states;
	hence, mode-independent definitions of entanglement were recently introduced \cite{SPBS19}.
	Furthermore, the problem of how---and if---the entanglement of particles and entanglement of fields are different is frequently met with a certain amount of vagueness (see, e.g., Refs. \cite{FG20,TWC91}), not only hindering theoretical progress but also limiting to what extent we can harness this resource.

	Another debate revolves around the exchange symmetry of matter particles and force-carrying fields that is sometimes considered to be one form of entanglement and sometimes considered a separate quantum effect.
	The nature of entanglement due to indistinguishability has been vastly studied (see, e.g., Refs. \cite{GMW02,BFT14,CCF18,BFFM20} and the references therein).
	Some works, however, have shown that the symmetrization of indistinguishable particles alone is a useful resource in quantum applications \cite{CMMTF07,KCP14,MYFZTA20,LPLKVAKTK18,FZDT18,KPSLFGGO18}.
	Despite the additional sophistication when it comes to telling symmetrization effects and other superposition phenomena apart \cite{SPBW17}, methods have been developed to characterize entanglement for symmetric and antisymmetric algebras \cite{ESBL02,GKM11,OK13,FLE13,RSV15}.

	In this contribution, we provide the framework to rigorously define and characterize entanglement between quantum particles and quantum fields.
	Specifically, we elaborate on the relation of quantum correlations to the quantum description in first (particles) and second quantization (fields).
	Instructive examples then prove that particle-entangled states can be field disentangled and vice versa,
	and both particle entanglement and field entanglement can be present and absent at the same time as well.
	Consequently, quantum correlations in particles and fields present two independent quantum effects---the aspect that we aim to call out here---that are both valid kinds of entanglement.
	In addition, for completeness, bosonic and fermionic systems are commented on, too, showing that the found independence applies regardless of exchange symmetry.

	We organize this paper as follows:
	A brief recapitulation of essential concepts and notations is provided in Sec. \ref{sec:Prelim}.
	We then formulate scenarios of quantum fields whose excitation yields distinguishable particles in Sec. \ref{sec:DistPart}.
	Here, we also put forward the independence of particle entanglemeent and field entanglement.
	In Sec. \ref{sec:InDistPart}, we proceed with some examples for bosons and fermions.
    We conclude in Sec. \ref{sec:Conclusion}.

\section{Preliminaries}
\label{sec:Prelim}

	In this section, we recapitulate concepts that are relevant in the context of this work.
	This includes establishing notations that are especially useful for identifying and relating first and second quantization.

\subsection{Fundamentals}

	For any number of subsystems, we can define the notion of a factorizable (pure) state, e.g.,
	\begin{equation}
		|\psi\rangle=|\psi_0\rangle\otimes|\psi_1\rangle\otimes|\psi_2\rangle\otimes\cdots.
	\end{equation}
	If the state of a composite quantum system is in this form, we can have local quantum interference but not global quantum correlations.
	Whenever the state of the system is not in this form, the parties are entangled, implying an insufficient description in terms of subsystem states alone.
	(For the sake of exposition, we focus on fully factorizable states throughout this work, and forms of partial entanglement can be deduced by combining multiple subsystems into one subsystem.)

	Beyond pure states, the concept of factorizability can be advanced to the notion of separable states \cite{W89}, allowing for classical correlation too.
	In this case, the mixed-state density operator in a separable (i.e., disentangled) configuration reads
	\begin{equation}
	    \label{eq:SepDef}
		\hat\rho=\int dP(\psi_0,\psi_1,\ldots)|\psi_0\rangle\langle\psi_0|\otimes |\psi_1\rangle\langle\psi_1| \otimes\cdots,
	\end{equation}
	where $P$ denotes a nonnegative joint probability distribution over pure tensor-product states.
	It is worth commenting that $P\ngeq 0$ in the above expansion allows one to decompose inseparable, i.e., entangled, states as well \cite{STV98}, as demonstrated in theory \cite{SW18} and experiment \cite{SMBBS19}.
	For simplicity, we study pure states {in this work};
	nevertheless, the {above} convex hull construction for {mixed} states can be used to elevate our approach from pure to mixed states straightforwardly.

	When talking about fermions and bosons, one has to respect the exchange symmetry when defining composite spaces \cite{MG64}.
	That is, rather than utilizing the tensor product $\otimes$, the symmetric tensor product $\vee$ and skew-symmetric (likewise, antisymmetric) tensor product $\wedge$ ought to be used, e.g.,
	\begin{equation}
		\label{eq:IndistTensorProd}
	\begin{aligned}
		|\psi_0\rangle\vee|\psi_1\rangle
		=&|\psi_0\rangle\otimes|\psi_1\rangle
		+|\psi_1\rangle\otimes|\psi_0\rangle
		\\
		\text{and}\quad
		|\psi_0\rangle\wedge|\psi_1\rangle
		=&|\psi_0\rangle\otimes|\psi_1\rangle
		-|\psi_1\rangle\otimes|\psi_0\rangle.
	\end{aligned}
	\end{equation}
	In the former case of bosons, we can also speak about two identical particles, $|\psi_0\rangle\vee|\psi_1\rangle$ with $|\psi_0\rangle=|\psi_1\rangle$, being a stronger restriction than just having indistinguishable particles.
	For the latter fermionic scenario, however, we have $|\psi_0\rangle\vee|\psi_1\rangle=0$ when $|\psi_0\rangle=|\psi_1\rangle$ holds true.
	Please be aware that $|\psi_0\rangle\parallel|\psi_1\rangle$ would suffice in both scenarios, but we ignore global phases and assume a proper normalization to one, $\langle\psi_0|\psi_0\rangle=1=\langle\psi_1|\psi_1\rangle$.

	In general, we include a normalization factor, $\mathcal N\in\mathbb C$, to indicate when a phase and normalization have been ignored.
	For example, $\mathcal N|\phi\rangle^{\vee n}$ represents a normalized and symmetric vector of $n$ identical bosons for $\mathcal N=(n!\langle\phi|\phi\rangle^n)^{-1/2}$.
	It is also worth reminding ourselves that $|\phi\rangle^{\otimes 0}$ is defined as the number one, similarly applying to $\wedge$ and $\vee$.

\subsection{Fock space}

	In first quantization, one considers a system with a fixed particle number.
	For instance, if the single-particle Hilbert space is $\mathcal H$, we have an $n$-particle space $\mathcal H^{\otimes n}$, assuming that all involved particles are distinguishable.
	In the cases of indistinguishable particles, one has $\mathcal H^{\vee n}$ and $\mathcal H^{\wedge n}$ for bosons and fermions, respectively.
	For the sake of simplicity, we describe all particles through the same Hilbert space $\mathcal H$. 
	In nature, one can observe interferences of states with different particle numbers too.
	For example, in principle, a laser yields a coherent superposition of $n$ identical photon states for all possible $n\in\mathbb N$, termed the coherent state.
	But, for example, a superposition of the form $\mathcal H^{\otimes n}\ni|\psi^{(n)}\rangle+|\psi^{(m)}\rangle\in\mathcal H^{\otimes m}$ for $m\neq n$ is ill-defined.
	The superscript, e.g., ``$(n)$'', indicates the number of particles in the state.

	The well-established Fock space \cite{F32}, related to second quantization, overcomes the aforementioned limitations, being defined through the direct sum of all $n$-particle spaces,
	\begin{equation}
	    \label{eq:FockSpace}
		\mathcal H_\mathrm{Fock}=\bigoplus_{n\in\mathbb N}\mathcal H^{\otimes n},
	\end{equation}
	where $\mathcal H^{\otimes 0}=\mathbb C$.
	(See Ref. \cite{SR07} for a thorough study of the transition from first to second quantization of light.)
	Because of this direct-sum-based definition, we can generally organize the fixed-particle components in a vector \cite{F32},
	\begin{equation}
	\begin{aligned}
		\begin{bmatrix}
			|\psi^{(0)}\rangle
			\\
			|\psi^{(1)}\rangle
			\\
			|\psi^{(2)}\rangle
			\\
			\vdots
		\end{bmatrix}
		=&
		\begin{bmatrix}
			|\psi^{(0)}\rangle
			\\
			0
			\\
			0
			\\
			\vdots
		\end{bmatrix}
		{+}
		\begin{bmatrix}
			0
			\\
			|\psi^{(1)}\rangle
			\\
			0
			\\
			\vdots
		\end{bmatrix}
		{+}
		\begin{bmatrix}
			0
			\\
			0
			\\
			|\psi^{(2)}\rangle
			\\
			\vdots
		\end{bmatrix}
		{+}\cdots
		\\
		=&\bigoplus_{n\in\mathbb N}|\psi^{(n)}\rangle=|\Psi\rangle_\mathrm{Fock},
	\end{aligned}
	\end{equation}
	providing the sought-after ability to interfere states with different particle numbers.
	The subscript ``Fock'' is used to distinguish states in Fock space from quantum states for a given {fixed} particle number.
	In addition, $\langle \psi^{(n)}|\psi^{(n)}\rangle$ is the probability to find $|\Psi\rangle_\mathrm{Fock}$ in the $n$-particle component.
	It is worth pointing out that $|\psi^{(0)}\rangle\in\mathcal H^{\otimes 0}=\mathbb C$ is a complex number {representing} vacuum ($|\psi^{(0)}\rangle=1$ {when} normalized), whose Fock-state representation is
	\begin{equation}
		|\mathrm{vac}\rangle_\mathrm{Fock}=
		\begin{bmatrix}
			1
			\\
			0
			\\
			0
			\\
			\vdots
		\end{bmatrix}.
	\end{equation}

	Furthermore, quantum field theories are typically formulated through annihilation and creation operators that act on the Fock space.
	We recapitulate that notion, also using the vector formalism described above.
	The generation of one extra (distinguishable) particle in the state $|\phi\rangle$ for each component of the Fock vector can be determined through the action
	\begin{equation}
	\begin{aligned}
		\hat a_\phi^\dag|\Psi\rangle_\mathrm{Fock}
		=&\bigoplus_{n\in\mathbb N}\left(|\phi\rangle\otimes|\psi^{(n)}\rangle\right)
		\\
		=&\begin{bmatrix}
			0 & 0 & 0 & \hdots
			\\
			|\phi\rangle\otimes & 0 & 0 & \hdots
			\\
			0 & |\phi\rangle\otimes & 0 & \ddots
			\\
			\vdots & \vdots & \ddots & \ddots
		\end{bmatrix}
		\begin{bmatrix}
			|\psi^{(0)}\rangle
			\\
			|\psi^{(1)}\rangle
			\\
			|\psi^{(2)}\rangle
			\\
			\vdots
		\end{bmatrix}
		\\
		=&\begin{bmatrix}
			0
			\\
			|\phi\rangle\otimes|\psi^{(0)}\rangle
			\\
			|\phi\rangle\otimes|\psi^{(1)}\rangle
			\\
			\vdots
		\end{bmatrix}.
	\end{aligned}
	\end{equation}
	As it is not relevant for our purpose, we ignore the commonly utilized factor $\sqrt{n+1}$ in the definition of the creation operator when it acts on the $n$-particle state.
	(Technically, our choice of factors relates to a so-called exponential phase operator \cite{SG64};
	a comparison of both kinds of factors can be found in Ref. \cite{MD02}.)
	The annihilation operator is obtained by Hermitian conjugation, $\hat a_\phi=(\hat a_\phi^\dag)^\dag$.
	Using our normalization, an $n$-fold excitation of the mode $\phi$ is given by
	\begin{equation}
		|n_\phi\rangle_\mathrm{Fock}=\hat a_\phi^{\dag n_\phi}|\mathrm{vac}\rangle_\mathrm{Fock},
	\end{equation}
	effectively resulting in a single, non-vanishing Fock-state component that is the $n_\phi$-fold tensor product of $|\phi\rangle$.
	Again, for indistinguishable particles, we replace $\otimes$ with the tensor products $\vee$ and $\wedge$ to obtain the desired exchange symmetry.
	For example, the aforementioned coherent state in the optical mode $\phi$ takes the general form
	\begin{equation}
		|\Psi\rangle_\mathrm{Fock}
		=\sum_{n\in\mathbb N} \Psi_{n} \hat a_\phi^{\dag n}|\mathrm{vac}\rangle_\mathrm{Fock}
		=\bigoplus_{n\in\mathbb N}\left(\Psi_{n}|\phi\rangle^{\vee n}\right).
	\end{equation}

	In the following, we apply all recapitulated facts {from this section} for the purpose of studying entanglement between particles and fields.
	To this end, we say that the single-particle Hilbert space $\mathcal H$ is given by the computational orthonormal basis
	\begin{equation}
		\label{eq:CompBasis}
		\{|j\rangle:j\in\mathbb N\}.
	\end{equation}
	Then, the second quantization for the $j$th mode is based on the creation operator $\hat a_j^\dag$,
	e.g., resulting in multimode number states
	\begin{equation}
		\label{eq:FockBasisState}
		|n_0,n_1,n_2,\ldots\rangle_\mathrm{Fock}=\hat a_0^{\dag n_0}\hat a_1^{\dag n_1}\hat a_2^{\dag n_2}\cdots|\mathrm{vac}\rangle_\mathrm{Fock},
	\end{equation}
	with the nonvanishing $(n_0+n_1+n_2+\cdots)$-particle component $|0\rangle^{\otimes n_0}\otimes|1\rangle^{\otimes n_1}\otimes|2\rangle^{\otimes n_2}\otimes\cdots$.

	It is worth pointing out that the order of the single-particle states---defining the modes---is determined through the order of action of the creation operators and may be changed as desired, e.g., using $\hat a_0^\dag\hat a_1^\dag\hat a_0^\dag$ to obtain $|0\rangle\otimes|1\rangle\otimes|0\rangle$.
	Nonetheless, our choice of ordering in Eq. \eqref{eq:FockBasisState} for distinguishable scenarios suffices for our intents and purposes,
	and the order becomes superfluous once we consider indistinguishability [see Eq. \eqref{eq:IndistTensorProd}].

\subsection{Outline}

	In the remainder of this work, our goal is to explore the uniting and distinct features of entanglement in the particle picture and for quantum fields.
	We provide explicit examples to show why field entanglement and particle entanglement are independent notions of quantum correlations.
	Along the way, rigorous definitions of particle entanglement and field entanglement are given within the Fock-vector-based framework, as described above.
	This allows us to clearly highlight the similarities and differences of both kinds of entanglement.

\section{Distinguishable particles}
\label{sec:DistPart}

	In this section, we establish the principles of particle entanglement and field entanglement.
	Specifically, we consider distinguishable particles and compare both forms of quantum correlations.
	Examples are constructed for all combinations, i.e., being only particle- or field-entangled, being disentangled with respect to either notion, and being simultaneously entangled in both forms.
	Thereby, we show the independence of field entanglement and particle entanglement.

\subsection{Particle entanglement}

	Suppose that we have $n$ particles, all of which are distinguishable.
	As discussed in the previous section, all particles are expanded in the computational basis [Eq. \eqref{eq:CompBasis}] for the sake of convenience.
	Now, by definition, we have an $n$-particle factorizable state $|\psi^{(n)}\rangle$ if we can write this state as the tensor product
	\begin{equation}
		\label{eq:ParticleFact}
		|\psi^{(n)}\rangle
		= |\psi_{1,n}\rangle\otimes\cdots\otimes|\psi_{n,n}\rangle.
	\end{equation}
	The labels in the subscript indicate the subsystem (particle label) and the total number of particles of the state, respectively.
	(Mixed-state separability for classically correlated systems follows the discussion in the previous section.)
	Conversely, we have entanglement in the $n$-particle state if
	$
		|\psi^{(n)}\rangle
		\neq |\psi_{1,n}\rangle\otimes\cdots\otimes|\psi_{n,n}\rangle
	$
	holds true.

	This approach can be generalized to the Fock-space description.
	That is, the state $|\Psi\rangle_\mathrm{Fock}$ is particle-factorizable if all $n$-particle components are $n$-particle factorizable,
	\begin{equation}
		\label{eq:ParticleFactFock}
		|\Psi\rangle_\mathrm{Fock}
		=\bigoplus_{n\in\mathbb N}\left(
			|\psi_{1,n}\rangle\otimes\cdots\otimes|\psi_{n,n}\rangle
		\right),
	\end{equation}
	i.e., each $n$-particle component is a tensor product of the form $|\psi_{1,n}\rangle\otimes\cdots\otimes|\psi_{n,n}\rangle$.
	For example, this approach was used to investigate macroscopic entanglement in systems with {fluctuating} numbers of particles \cite{SW17}.

\subsection{Field entanglement}

	Using the modes that are determined by our computational basis [Eq. \eqref{eq:CompBasis}], we can now define field factorizability through the relation
	\begin{equation}
		\label{eq:FieldFact}
	\begin{aligned}
	|\Psi\rangle_\mathrm{Fock}
		=&
		|\Psi_0\rangle_\mathrm{Fock}
		\circledast|\Psi_1\rangle_\mathrm{Fock}
		\circledast|\Psi_2\rangle_\mathrm{Fock}
		\circledast\cdots.
	\end{aligned}
	\end{equation}
	Herein, $\circledast$ indicates the standard tensor product of Fock spaces;
	this symbol is used here for the purpose of distinguishing it from the tensor product $\otimes$ as used previously for fixed particle numbers.
	With $\circledast$, Fock basis elements are
	\begin{equation}
		\label{eq:FockTensorProduct}
		|n_0,n_1,\ldots\rangle_\mathrm{Fock}
		{=}|n_0\rangle_\mathrm{Fock}\circledast
		|n_1\rangle_\mathrm{Fock}\circledast
		{\cdots}
	\end{equation}
	for $n_0,n_1,\ldots\in\mathbb N$.
	See also Eq. \eqref{eq:FockBasisState} in this context.
	Being defined for the basis elements, this tensor product can thereby be extended to arbitrary states in Eq. \eqref{eq:FieldFact}, using the expansion
	\begin{equation}
		\label{eq:FockElementExpansion}
		|\Psi_j\rangle_\mathrm{Fock}=\sum_{n_j\in\mathbb N}\Psi_{n_j}|n_j\rangle_\mathrm{Fock}
		\quad\text{for}\quad
		j\in\mathbb N.
	\end{equation}
	By definition, if Eq. \eqref{eq:FieldFact} is falsified, the state under study is field-entangled (also commonly referred to as mode-entangled).

	For a preliminary comparison with particle entanglement, we expand the state in Eq. \eqref{eq:FieldFact} in the Fock basis, additionally utilizing Eq. \eqref{eq:FockElementExpansion}, according to its $N$-particle components.
	We find
	\begin{equation}
		\label{eq:FieldFactExpansion}
	\begin{aligned}
		&
		|\Psi_0\rangle_\mathrm{Fock}
		\circledast|\Psi_1\rangle_\mathrm{Fock}
		\circledast\cdots
		\\
		=&
		\sum_{n_0,n_1,\ldots\in\mathbb N}\Psi_{n_0,0}\Psi_{n_1,1}\cdots|n_0,n_1,\ldots\rangle_\mathrm{Fock}
		\\
		=&
		\bigoplus_{N\in\mathbb N}\left(
			\sum_{\substack{ n_0,n_1,\ldots\in\mathbb N: \\ n_0+n_1+\cdots=N}}
			\prod_{j\in\mathbb N}
			\Psi_{n_j,j}\,
			|0\rangle^{\otimes n_0}\otimes|1\rangle^{\otimes n_1}\otimes\cdots
		\right).
	\end{aligned}
	\end{equation}
	Note that the $N$-particle component takes the form of a linear combination, i.e., superpositions.

	The similarities between particle entanglement and field entanglement are clear.
	Either kind of quantum correlation violates a tensor-product form, Eqs. \eqref{eq:ParticleFact} and \eqref{eq:FieldFact}, applying tensor products $\otimes$ and $\circledast$ which respectively operate on the multiparticle Hilbert space and the Fock space.
	Thus, their mathematical structures are similar.
	On the physical side, both kinds of correlations answer the question of whether a state can be fully described by local quantum states using either individual particles or distinct modes.
	The examples we construct in the following, however, show the dissimilarities between particle-entangled and field-entangled states, demonstrating their inequivalence.

\subsection{Proof-of-concept examples}

	In our examples here, we mainly focus on the qubit space (i.e., having two distinct modes, the $0$th and the $1$st) for the sake of exposition.
	We start with the Fock states as previously defined, which are \textit{both field- and particle-factorizable},
	\begin{equation}
	\begin{aligned}
		&|n_0\rangle_\mathrm{Fock}\circledast|n_1\rangle_\mathrm{Fock}
		=\hat a_0^{\dag n_0}\hat a_1^{\dag n_1}|\mathrm{vac}\rangle_\mathrm{Fock}
		\\
		=& \bigoplus_{N\in\mathbb N}\left(
		\delta_{N,n_0+n_1} |0\rangle^{\otimes n_0}\otimes|1\rangle^{\otimes n_1}
		\right),
	\end{aligned}
	\end{equation}
	with the Kronecker symbol $\delta_{N,n_0+n_1}=1$ for $N=n_0+n_1$ and zero otherwise.
	That is, these states are tensor products with respect to $\circledast$ and all $N$-particle components, $|0\rangle\otimes\cdots\otimes|0\rangle\otimes|1\rangle\otimes\cdots\otimes|1\rangle$.
	Keep in mind that the algebraic null $0$ for $n_0+n_1\neq N$ can always be seen as an arbitrary product state with a zero probability amplitude in the Fock representation.
	In addition, it is noteworthy that many more states which are factorizable with respect to particles and fields exist.
	Our objective is only to provide at least one representative example for all the different kinds of quantum correlations within the classification of particle entanglement and field entanglement.

	Our second example concerns a \textit{field-entangled and particle-factorizable state}.
	The two-mode state that exemplifies that is
	\begin{equation}
	\begin{aligned}
		&\mathcal N\sum_{n\in\mathbb N}\lambda^{n} |n,n\rangle_\mathrm{Fock}
		\\
		=&\mathcal N\bigoplus_{N\in\mathbb N: N=2n}
		\lambda^{n}\left(|0\rangle^{\otimes n}\otimes|1\rangle^{\otimes n}\right),
	\end{aligned}
	\end{equation}
	using the normalization constant $\mathcal N$ and a complex parameter $\lambda$, with $0<|\lambda|<1$.
	In optics, this state relates to a so-called two-mode squeezed (vacuum) state.
	In the Fock expansion, we have a Schmidt decomposition that has an infinite (specifically, larger than one) Schmidt rank \cite{NC10}, certifying entanglement.
	Each $N$-particle component, however, is clearly a product state of the form $|0\rangle\otimes\cdots\otimes|0\rangle\otimes|1\rangle\otimes\cdots\otimes|1\rangle$ and is thus factorizable for all particle numbers.

	In contrast to the previous example, we can also have \textit{field-factorizable and particle-entangled states}.
	A representative is
	\begin{equation}
		\label{eq:FieldSepPartInsep}
	\begin{aligned}
		&\mathcal N\left(|\Phi\rangle_\mathrm{Fock}\circledast|\Phi\rangle_\mathrm{Fock}\right)
		\\
		=&\mathcal N\bigoplus_{N\in\mathbb N}\left(
			\sum_{\substack{n,m\in\mathbb N:\\ n+m=N}}
			\lambda^n\lambda^{m}\,
			|0\rangle^{\otimes n}\otimes|1\rangle^{\otimes m}
		\right)
		\\
		=&\mathcal N\bigoplus_{N\in\mathbb N}\lambda^N\left(
			|1\rangle\otimes|1\rangle^{\otimes (N-1)}
			+|0\rangle\otimes|\phi^{(N-1)}\rangle
		\right).
	\end{aligned}
	\end{equation}
	Therein, both modes are individually given through $|\Phi\rangle_\mathrm{Fock}=\sum_{n\in\mathbb N}\lambda^n|n\rangle_\mathrm{Fock}$.
	By contrast, the $N$-particle component takes a superposition form (see the second line), as discussed in the context of Eq. \eqref{eq:FieldFactExpansion}.
	In the third line of Eq. \eqref{eq:FieldSepPartInsep}, a factorization of the components for $N>1$ fails because the decomposition includes the linearly independent parts $|1\rangle^{\otimes (N-1)}$ and $|\phi^{(N-1)}\rangle=\sum_{n\in\mathbb N:1\leq n\leq N}|0\rangle^{\otimes n}\otimes|1\rangle^{\otimes (N-1-n)}$, implying a Schmidt rank of two, like a Bell state.
	Particle entanglement of this field-factorizable state is thus proven.
	
    It is worth emphasizing that the previous two examples that are either field- or particle-entangled depend on different superpositions.
    That is, the particle entanglement was demonstrated by superimposing states in first quantization, while field entanglement was achieved by superpositions in second-quantization Fock space.
    Together with the products $\otimes$ and $\circledast$, this led to mutually exclusive forms of quantum correlations.
    
	For completeness, we can also consider a \textit{field-entangled and particle-entangled state}.
	In particular, we construct an example whose particle entanglement corresponds to GHZ states and field quantum correlations resemble a so-called NOON state \cite{KLD02}.
	The state has the form
	\begin{equation}
	\begin{aligned}
		&\mathcal N\left(
			|\Phi\rangle_\mathrm{Fock}\circledast|0\rangle_\mathrm{Fock}
			{+}|0\rangle_\mathrm{Fock}\circledast|\Phi\rangle_\mathrm{Fock}
		\right)
		\\
		=&\mathcal N\sum_{N\in\mathbb N}
		\lambda^N\left(|N,0\rangle_\mathrm{Fock}+|0,N\rangle_\mathrm{Fock}\right)
		\\
		=&\mathcal N\bigoplus_{N\in\mathbb N}\lambda^N\left(
			|0\rangle^{\otimes N}+|1\rangle^{\otimes N}
		\right),
	\end{aligned}
	\end{equation}
	using the fact that $|\Phi\rangle_\mathrm{Fock}$, as given before, is linearly independent of $|0\rangle_\mathrm{Fock}$.
	As we have particle entanglement and field entanglement at the same time, we could also speak about a form of hybrid entanglement here that allows harnessing the properties of GHZ states in the particle picture and the capabilities of NOON states when operating on quantized fields.

\subsection{Preliminary summary}

	In summary, we have demonstrated that particle entanglement and field entanglement are phenomena which can occur independently.
	In other words, particle entanglement does not imply field entanglement, nor does field entanglement imply particle entanglement.
	This extends to mixed states [Eq. \eqref{eq:SepDef}] and notions of field and particle separability and inseparability as well.
	Since entanglement serves as a resource in many applications, this also means that particle entanglement and field entanglement are resources that are independently accessible, allowing for high flexibility in diverse quantum technology applications that can simultaneously harness field and particle quantum correlations.

\section{Indistinguishable particles}
\label{sec:InDistPart}
    
    For completeness, we also comment on the case of indistinguishable systems.
    As pointed out before, quantum correlations in such scenarios have been widely discussed in terms of theoretical and experimental developments.
    See, e.g., Ref. \cite{BFFM20} for a thorough classification.
    While it is not our intention to dive into all the finer details of entanglement of indistinguishable particles, we still want to show that field entanglement and particle entanglement are independent phenomena in such scenarios too.
    To this end, below, we lay out the adopted notation and provide examples for all combinations of factorizability and entanglement of indistinguishable particles and fields, thus completing our catalog of examples of the independence of field entanglement and particle entanglement for all possible scenarios.

\subsection{Bosonic quantum fields}
\label{sec:InDistPartBoson}

	For indistinguishable particles, such as bosons, the notion of particle entanglement is contentious.
	This is due to the definition of the symmetric tensor product $\vee$ [Eq. \eqref{eq:IndistTensorProd}], which generally produces a nonlocal superposition with respect to $\otimes$.
	This leads to two possible definitions of separability.
	In the community, there have been discussions from defenders of both definitions \cite{GMW02,BFT14,CCF18}, and both sides have {produced} convincing arguments, in particular, when focusing on applications \cite{CMMTF07,KCP14,MYFZTA20}.
	An agnostic standpoint is taken here, meaning that we discuss all options of particle entanglement in bosonic {(and later in fermionic)} systems without any preference, and the reader is invited to select their preferred point of view.

	We begin with two bosons to discuss different options of factorizability.
	For $n>2$ bosons, the general ideas presented here can be generalized.
	We compare the two provided factorizations through a convenient decomposition and present an example of entanglement that is applicable to both notions.

	First, one can say that a system of two indistinguishable particles with symmetric exchange symmetry is particle-factorizable if
	\begin{equation}
	\begin{aligned}
		|\psi^{(2)}\rangle
		=&\mathcal N|\psi_1\rangle\vee|\psi_2\rangle
		\\
		=&\mathcal N\big(
			|\psi_1\rangle\otimes|\psi_2\rangle
			+|\psi_2\rangle\otimes|\psi_1\rangle
		\big),
	\end{aligned}
	\end{equation}
	for any $|\psi_1\rangle,|\psi_2\rangle\in\mathcal H$.
	Second, when additionally requiring identical (states of) particles, we have particle factorizability that reads
	\begin{equation}
		|\psi^{\prime (2)}\rangle=\mathcal N |\psi^{(1)}\rangle\vee|\psi^{(1)}\rangle=2\mathcal N|\psi^{(1)}\rangle\otimes|\psi^{(1)}\rangle
	\end{equation}
	for arbitrary one-particle states $|\psi^{(1)}\rangle\in\mathcal H$, which is also factorizable with respect to $\otimes$.
	One can readily see that factorizability in the form $|\psi^{\prime (2)}\rangle$ implies factorizability as described through $|\psi^{(2)}\rangle$ with $|\psi_1\rangle=|\psi_2\rangle=|\psi^{(1)}\rangle$.
	Conversely, entanglement of indistinguishable bosons yields entanglement of identical bosons.

	For an additional comparison, we decompose the state $|\psi^{(2)}\rangle$ in terms of states of the form $|\psi^{\prime (2)}\rangle$.
	Here, it is convenient to define the following parameter and states:
	\begin{equation}
		\label{eq:ForBosonDecomp}
		\gamma=\langle\psi_1|\psi_2\rangle
		\quad\text{and}
		\quad
		|e_\pm\rangle=\frac{
			|\psi_1\rangle
			\pm\frac{\gamma^\ast}{|\gamma|}
			|\psi_2\rangle
		}{\sqrt{2(1\pm|\gamma|)}},
	\end{equation}
	where $|e_+\rangle$ and $|e_-\rangle$ are orthonormal.
	Note that we suppose that $\langle\psi_1|\psi_1\rangle=1=\langle\psi_2|\psi_2\rangle$ is obeyed and that $|\psi_1\rangle$ and $|\psi_2\rangle$ are neither parallel nor orthogonal to exclude trivial cases.
	By using the quantities in Eq. \eqref{eq:ForBosonDecomp}, we obtain
	\begin{equation}
	\begin{aligned}
		|\psi^{(2)}\rangle
		=&\frac{
			|\psi_1\rangle\otimes|\psi_2\rangle
			+|\psi_2\rangle\otimes|\psi_1\rangle
		}{
			\sqrt{2(1+|\gamma|^2)}
		}
		\\
		=&
		\frac{\gamma}{|\gamma|}\frac{1+|\gamma|}{\sqrt{2(1+|\gamma|^2)}}
		|e_+\rangle\otimes|e_+\rangle
		\\
		&-\frac{\gamma}{|\gamma|}\frac{1-|\gamma|}{\sqrt{2(1+|\gamma|^2)}}
		|e_-\rangle\otimes|e_-\rangle.
	\end{aligned}
	\end{equation}
	This shows that $|\psi^{(2)}\rangle$ is a superposition of factorizable states of identical particles, $|e_+\rangle\vee|e_+\rangle$ and $|e_-\rangle\vee|e_-\rangle$, while being factorizable for distinguishable particles, $|\psi_1\rangle\vee|\psi_2\rangle$.

	An example for a state that is entangled with respect to both notions of factorizable bosons is
	\begin{equation}
		\label{eq:chiStateTwoBosons}
		|\chi^{(2)}\rangle=\mathcal N\left(|0\rangle\otimes|1\rangle+|1\rangle\otimes|0\rangle+|2\rangle\otimes|2\rangle\right).
	\end{equation}
	Applying the above results, this state is a superposition of the following products of three identical boson states:
	$(|0\rangle+|1\rangle)^{\vee 2}$, $(|0\rangle-|1\rangle)^{\vee 2}$, and $|2\rangle^{\vee 2}$.
	At the same time, it is a superposition of the two symmetric product states $|0\rangle\vee|1\rangle$ and $|2\rangle^{\vee 2}$ of indistinguishable bosons.

	Regardless which of the aforementioned definitions is used for bosons, we have particle factorizability in the Fock representation when
	\begin{equation}
		|\Psi\rangle_\mathrm{Fock}=\bigoplus_{n\in\mathbb N}\left(|\psi_{1,n}\rangle\vee\cdots\vee|\psi_{n,n}\rangle\right)
	\end{equation}
	applies, where we additionally require that $|\psi_{1,n}\rangle=\cdots=|\psi_{n,n}\rangle$ in the case of products of identical particles.
	For correspondingly entangled states of bosons, the above representation does not apply. 
	Except for the details pertaining to the exchange symmetry, which we discussed already, this definition of particle factorizability is analogous to the definition of distinguishable particles in Eq. \eqref{eq:ParticleFactFock}.
    Also, because of the symmetry, the order of creation operators no longer plays a role, and all orderings we choose for distinguishable particles become superfluous.

	The Fock basis for computational modes is thus given by
	\begin{equation}
	\begin{aligned}
		&|n_0,n_1,\ldots\rangle_\mathrm{Fock}
		=\mathcal N\hat a_0^{\dag n_0}\hat a_1^{\dag n_1}\cdots |\mathrm{vac}\rangle_\mathrm{Fock}
		\\
		=&\mathcal N\bigoplus_{N\in\mathbb N}\left(
			\delta_{N,n_0+n_1+\cdots}\,
			|0\rangle^{\vee n_0}\vee
			|1\rangle^{\vee n_1}\vee
			\cdots
		\right).
	\end{aligned}
	\end{equation}
	Interestingly, however, formally, the same tensor-product over Fock spaces applies as in the case of distinguishable particles [see Eq. \eqref{eq:FockTensorProduct}] to define product bases in Fock space, e.g., $|n_0,n_1\rangle_\mathrm{Fock}=|n_0\rangle_\mathrm{Fock}\circledast|n_1\rangle_\mathrm{Fock}$,
	for the zeroth and first modes.
	Moreover, the same applies to the actual definition of field-factorizable states in bosonic systems, for which we can directly copy the definition from the previous section [see Eq. \eqref{eq:FieldFact}].
	Therefore, the mode-based notion of entanglement has not fundamentally changed compared to the earlier scenario of distinguishable particles.

	Again, we can construct examples to highlight the distinct features of particle entanglement and field entanglement.
	As before, Fock basis states are field factorizable, e.g., $|2,0\rangle_\mathrm{Fock}$ and $|1,1\rangle_\mathrm{Fock}$ for modes $0$ and $1$.
	The state $|2,0\rangle_\mathrm{Fock}$ is also factorizable with respect to indistinguishable and identical bosons, resembling $|0\rangle^{\vee 2}$, and the state $|1,1\rangle_\mathrm{Fock}$ is factorizable with respect to indistinguishable bosons, $|0\rangle\vee|1\rangle$, but entangled for identical bosons, as discussed via the decomposition of two-boson states above.

	A state that is \textit{field-entangled and factorizable for indistinguishable and identical bosons} is
	\begin{equation}
	\begin{aligned}
		|\Psi'\rangle_\mathrm{Fock}
		=&\mathcal N\begin{bmatrix}
			0 \\ 0 \\ |0\rangle^{\vee 2} \\ |1\rangle^{\vee 3} \\ 0 \\ \vdots
		\end{bmatrix}
		\\
		=&\frac{|2,0\rangle_\mathrm{Fock}+3|0,3\rangle_\mathrm{Fock}}{\sqrt{10}}.
	\end{aligned}
	\end{equation}
	By construction, this state is factorizable in the particle picture and entangled in terms of field components, as deduced from the Schmidt number of two in Fock space.
	The simple modification
	\begin{equation}
	\begin{aligned}
		|\Psi\rangle_\mathrm{Fock}
		=&\mathcal N\begin{bmatrix}
			0 \\ 0 \\ |0\rangle\vee|1\rangle \\ |1\rangle^{\vee 3} \\ 0 \\ \vdots
		\end{bmatrix}
		\\
		=&\frac{|1,1\rangle_\mathrm{Fock}+3\sqrt{2}|0,3\rangle_\mathrm{Fock}}{\sqrt{19}}.
	\end{aligned}
	\end{equation}
	is rather similar, except that we now \textit{loose factorizability with respect to identical particles} because of the modified two-particle component because $|0\rangle\vee|1\rangle\neq |\psi^{(1)}\rangle^{\vee 2}$.

	Conversely, states that are \textit{field-factorizable and particle-entangled} can be constructed too.
	For example, the product vector $(|0\rangle_\mathrm{Fock}+|1\rangle_\mathrm{Fock})^{\circledast 3}$ has a two-particle component proportional to $(|0\rangle\vee|1\rangle+|1\rangle\vee|2\rangle+|2\rangle\vee|0\rangle)$.
	In a few steps, one can show that this is not factorizable for indistinguishable particles, implying the same for identical bosons:
	equating $|0\rangle\vee|1\rangle+|2\rangle\vee|2\rangle+|2\rangle\vee|0\rangle$ with the ansatz $(\alpha_0|0\rangle+\alpha_1|1\rangle+\alpha_2|2\rangle) \vee (\beta_0|0\rangle+\beta_1|1\rangle+\beta_2|2\rangle)$ means that we have to set, without loss of generality, $\alpha_0=0$ so that we do not have a $|0\rangle\vee|0\rangle$ component, which implies that $\alpha_1=\alpha_2=1/\beta_0$; 
	this, in turn, implies that $\beta_1=\beta_2=0$ so that the $|1\rangle\vee|1\rangle$ and $|2\rangle\vee|2\rangle$ components are removed;
	however, now, the $|1\rangle\vee|2\rangle$ component also comes with a zero factor, contradicting the decomposition of the given two-particle state.

	We can also find a state that is entangled with respect to all three notions that are investigated in this section.
	For example, the following three-mode state, using Eq. \eqref{eq:chiStateTwoBosons}, satisfies our demands:
	\begin{equation}
	\begin{aligned}
		|X\rangle_\mathrm{Fock}
		=&\begin{bmatrix}
			0 \\ 0 \\ |\chi^{(2)}\rangle \\ 0 \\ \vdots
		\end{bmatrix}
		\\
		=&\frac{\sqrt{2}|1,1,0\rangle_\mathrm{Fock}+|0,0,2\rangle_\mathrm{Fock}}{\sqrt{3}}.
	\end{aligned}
	\end{equation}
	As described above, $|\chi^{(2)}\rangle$ [Eq. \eqref{eq:chiStateTwoBosons}] is particle-entangled with respect to both discussed options.
	Furthermore, factoring the zeroth mode already fails since the remaining two-mode vectors $|1,0\rangle_\mathrm{Fock}$ and $|0,2\rangle_\mathrm{Fock}$ are linearly independent, resulting in a Schmidt rank larger than one.

	To summarize, we again find that field entanglement and particle entanglement are independent phenomena for bosons.
	Examples were constructed that showed that quantum correlations between fields and identical particles can occur independently, which holds true for indistinguishable particles as well.
    However, the two forms of particle entanglement for which there is no case of factorizability in terms of identical states and entanglement for symmetrized (i.e., only indistinguishable) states are dependent on each other,, as discussed at the beginning of the section.
	The definition of entanglement between modes is not conceptually different from the case without exchange symmetry.
	Since the exchange symmetry acts on the level of individual particles, this observation makes sense.

\subsection{Indistinguishable particles: Quantum fields of fermions}
\label{sec:InDistPartFermion}

	Last, we complete our investigation of entanglement between fields and particles by studying scenarios with skew-symmetric fermions.
	In contrast to the bosonic particles, where $|\psi\rangle\otimes|\psi\rangle=\mathcal N|\psi\rangle\vee|\psi\rangle$ presents a standard tensor product as well as a symmetric tensor product, the Pauli exclusion principle, mathematically determined through $|\psi\rangle\wedge|\psi\rangle=0$, does not allow for factorizability of identical fermions.
	Thus, from a purist perspective, all fermionic states, including examples like $|0\rangle\wedge|1\rangle=|0\rangle\otimes|1\rangle-|1\rangle\otimes|0\rangle$, could be considered to be particle-entangled.
	This renders it trivial to study entanglement of identical particles.
	Therefore, we here focus on the notion that states of the form $|\psi_1\rangle\wedge\cdots\wedge|\psi_n\rangle$ are factorizable with respect to the antisymmetric tensor product, i.e., the distinguishable case.
	
    Furthermore, antisymmetry further implies that each basis state $|j\rangle$ can be occupied once at most.
    Specifically, we have $\hat a_j^{\dag 2}=0$ because it acts like the operator $|j\rangle\wedge|j\rangle\wedge=0$ on each $n$-particle component of the Fock space, as discussed in Sec. \ref{sec:Prelim}.
    Thus, Fock basis elements take a form considered previously, $|n_0\rangle_\mathrm{Fock}\circledast|n_1\rangle_\mathrm{Fock}\circledast\cdots$, but with the significant restriction $n_j\in\{0,1\}$ for each mode $j$, rather than $n_j\in\mathbb N$.

    As a final remark, we mention that the Schmidt decomposition (also known as singular-value decomposition) that is applicable to distinguishable particles and can be adjusted to bosons (Takagi's factorization) is replaced by the so-called Slater decomposition for fermions.
    For instance, using the definitions in Eq. \eqref{eq:ForBosonDecomp}, we have
    \begin{equation}
    \begin{aligned}
        |\psi^{(2)}\rangle
        =&\mathcal N |\psi_1\rangle\wedge|\psi_2\rangle
        \\
        =&\frac{|\psi_1\rangle\otimes|\psi_2\rangle-|\psi_2\rangle\otimes|\psi_1\rangle}{\sqrt{2(1-|\gamma|^2)}}
        \\
        =&\frac{1}{\sqrt2}\frac{\gamma}{|\gamma|}\left(
            |e_-\rangle\otimes|e_+\rangle-|e_+\rangle\otimes|e_-\rangle
        \right).
    \end{aligned}
    \end{equation}
    Hence, this state is proportional to the skew-symmetric product of two orthonormal states, $|e_-\rangle\wedge|e_+\rangle$, and has a so-called Slater rank of one \cite{ESBL02,RSV15}.
    Through this unit rank, we can thus identify factorizability for skew-symmetric tensors analogously to a Schmidt rank of one for distinguishable particles.

    As done previously for distinguishable particles and indistinguishable and identical bosons, we now provide examples to display the independence of field entanglement and particle entanglement for systems of indistinguishable fermions.
    For instance, this is relevant for modern research at the interface of quantum chemistry and quantum information \cite{DS20,DMDSSZS21}.

    Very much like in the case of distinguishable particles and bosonic fields, we find that the basis states of the Fock space are field-factorizable and particle-factorizable at the same time.
    That is, the $(M+1)$-mode Fock basis states $|n_0\rangle_\mathrm{Fock}\circledast\cdots\circledast|n_M\rangle_\mathrm{Fock}$ are a product with respect to $\circledast$, and the only nonvanishing particle component of the Fock vector, which the one pertaining to $n_0+\cdots+n_M$ particles, reads $|0\rangle^{\wedge n_0}\wedge\cdots\wedge|M\rangle^{\wedge n_M}$.

    On the other hand, we can easily find examples in which we have particle entanglement and field entanglement at the same time.
    For instance, we can consider a generalized GHZ-type state in the particle picture, proportional to
    \begin{equation}
    \begin{aligned}
        &|0\rangle\wedge\cdots\wedge|M-1\rangle+|M\rangle\wedge\cdots\wedge|2M-1\rangle
        \\
        &+\cdots+|(L-1)M\rangle\wedge\cdots\wedge|LM-1\rangle,
    \end{aligned}
    \end{equation}
    which superimposes $LM$ occupied modes, where $L,M\in\mathbb N$.
    In the field notation of Fock spaces, the same state resembles a W-type state,
    \begin{equation}
    \begin{aligned}
        &|1\rangle_\mathrm{Fock}^{\circledast M}\circledast|0\rangle_\mathrm{Fock}^{\circledast (L-1)M}
        \\
        &+|0\rangle_\mathrm{Fock}^{\circledast M}\circledast|1\rangle_\mathrm{Fock}^{\circledast M}\circledast|0\rangle_\mathrm{Fock}^{\circledast (L-2)M}
        \\
        &+\cdots+|0\rangle_\mathrm{Fock}^{\circledast (L-1)M}\circledast|1\rangle_\mathrm{Fock}^{\circledast M}.
    \end{aligned}
    \end{equation}
    Therefore, this fermionic state is entangled in the particle and field sense.

    Finally, examples for particle-factorizable and field-entangled state and vice versa are provided.
    The two-particle state $\mathcal N(|0\rangle+|1\rangle)\wedge(|2\rangle+|3\rangle)$ reads like a two-fold copy of entangled Bell states in the Fock representation, $\mathcal N(|0,1\rangle_\mathrm{Fock}+|1,0\rangle_\mathrm{Fock})^{\circledast 2}$.
    Hence, we simultaneously have a particle-factorizable and field-entangled state.
    On the other hand, a particle-entangled and field-factorizable state is $\mathcal N(|0\rangle_\mathrm{Fock}+|1\rangle_\mathrm{Fock})^{\circledast 4}$.
    This state has a two-particle component proportional to
    \begin{equation}
    \begin{aligned}
        &
        |0\rangle\wedge|1\rangle
        +|0\rangle\wedge|2\rangle
        +|0\rangle\wedge|3\rangle
        \\
        &+|1\rangle\wedge|2\rangle
        +|1\rangle\wedge|3\rangle
        +|2\rangle\wedge|3\rangle
        \\
        =&(|0\rangle+|1\rangle)\wedge(|1\rangle+|2\rangle+|3\rangle)+|2\rangle\wedge|3\rangle,
    \end{aligned}
    \end{equation}
    where the latter form is a sum of two products of linearly independent vectors.
    This implies a Slater rank (of two) greater than one and thus nonfactorizability with respect to $\wedge$.

    In conclusion, for all combinations of factorizability and entanglement of fermionic particles and fields, we have provided examples.
    Th{e resulting} independence of particle entanglement and field entanglement for fermions completes our analysis of distinguishable and indistinguishable particles.

\section{Conclusion}
\label{sec:Conclusion}

    We investigated different notions of entanglement as a consequence of first and second quantization in quantum physics, providing a framework to rigorously define and characterize entanglement between particles, fields, and their hybrid combinations.
    We started by establishing distinct notations and highlighting the role of Fock space to precisely define the different notions of separability.
    Examples were explicitly analyzed, {extending to} cases for distinguishable and indistinguishable particles.
    Importantly, it was shown that quantum correlations in particles and fields are independent quantum effects since particle-entangled states can be field-disentangled and field-entangled states can be particle-disentangled.
    In addition, we presented examples of full separability with respect to either notion and joint particle-field entanglement.
    Therefore, the quite common assumption that entanglement of fields implies entanglement of particles or vice versa is rendered obsolete.
    Rather, the findings of this work demonstrate on the basis of concrete examples that it does matter for the fundamental notion of entanglement whether one studies it in the context of first or second quantization.
    As the exchange of symmetry could play a role, we also showed that the same concept of independence applies to bosonic and fermionic systems.

    The independent kinds of entanglement were rigorously determined here with the aim of deepening fundamental understanding of essential quantum correlations.
    In addition, our investigation has direct implications for practical quantum technologies;
    in particular, we showed that particle entanglement and field entanglement provide two distinct resources for quantum information processing that can be exploited separately and jointly.
    For instance, states that are both particle- and field-entangled may serve as an interface to convert field-based quantum resources to particle-based technology platforms and vice versa.

\begin{acknowledgments}
	J.S. acknowledges financial support from the Deutsche Forschungsgemeinschaft (DFG, German Research Foundation) through the Collaborative Research Center TRR 142 (Project No. 231447078, project C10).
	E.A. acknowledges funding from the Der Wissenschaftsfonds FWF (Fonds zur F\"{o}rderung der wissenschaftlichen Forschung) Lise Meitner-Programm (M3151).
\end{acknowledgments}


\end{document}